\documentclass[aip,jcp,graphicx,reprint]{revtex4-1}
\usepackage{graphicx} 
\usepackage{booktabs}
\usepackage{multirow}
\usepackage{amsmath}

\begin{document}

\title{A Multimer Embedding Approach for Molecular Crystals up to Harmonic Vibrational Properties} 

\author{Johannes Hoja}
\email[]{johannes.hoja@uni-graz.at}

\author{Alexander List}

\author{A. Daniel Boese}
\email[]{adrian\_daniel.boese@uni-graz.at}

\affiliation{Department of Chemistry, University of Graz, Heinrichstraße 28/IV, 8010 Graz, Austria.}

\date{\today}

\begin{abstract}
Accurate calculations of molecular crystals are crucial for drug design and crystal engineering. However, periodic high-level density functional calculations using hybrid functionals are often prohibitively expensive for relevant systems. 
These expensive periodic calculations can be circumvented by the usage of embedding methods in which for instance the periodic calculation is only performed at a lower-cost level and then monomer energies and dimer interactions are replaced by those of the higher-level method. 
Herein, we extend upon such a multimer embedding approach to enable energy corrections for trimer interactions and the calculation of harmonic vibrational properties up to the dimer level. 
We evaluate this approach for the X23 benchmark set of molecular crystals by approximating a periodic hybrid density functional (PBE0+MBD) by embedding multimers into less expensive calculations using a generalized-gradient approximation (GGA) functional (PBE+MBD). 
We show that trimer interactions are crucial for accurately approximating lattice energies within 1 kJ/mol and might also be needed for further improvement of lattice constants and hence cell volumes. 
Finally, vibrational properties are already very well captured at the monomer and dimer level, making it possible to approximate vibrational free energies at room temperature within 1 kJ/mol.
\end{abstract}


\maketitle 

\section{Introduction}
The capability to accurately but still efficiently model molecular crystals would be invaluable for crystal engineering\cite{Desiraju2013} and drug design\cite{Datta2004}. 
However, the individual molecules within molecular crystals are only weakly held together by non-covalent interactions and for many molecules, different crystal-packing arrangements are possible.
Such different polymorphs can have very similar lattice energies\cite{CruzCabeza2015}, which often differ by only a few kJ/mol.
Therefore, it is vital to accurately capture the subtle interplay of intermolecular interactions. 
Furthermore, single-point energy calculations or simple lattice relaxations are also often insufficient since many properties of molecular crystals can highly depend on temperature and pressure\cite{Hoja2016}. 
For instance, the actual relative stability of polymorphs can often not be determined by static lattice energies alone, but rather free energies have to be considered\cite{Reilly2014,Nyman2015,Hoja2019}, which means that computationally expensive vibrational free energies have to be calculated as well. 
In addition, it might often also be necessary to explicitly account for the thermal expansion of the crystal\cite{Dolgonos2019,Hoja2016,Heit2016,McKinley2018,Brandenburg2017}, which requires harmonic phonon calculations for several different unit-cell volumes.

Given the periodic nature and the often large unit-cell sizes of practically relevant molecular crystals, highly accurate wave-function methods cannot routinely be used and we have to rely on more approximate methods. 
Currently, the main workhorse for high-level calculations of molecular crystals is periodic density functional theory (DFT). 
One important way of assessing the quality of computational methods under realistic conditions are the regular crystal structure prediction blind tests organized by the Cambridge Crystallographic Data Centre\cite{Lommerse2000,Motherwell2002,Day2005,Day2009,Bardwell2011,Reilly2016}, with the current blind test just having completed in 2022. 
Therein, van-der-Waals dispersion inclusive density functional approximations are often very successfully used in the final steps of such crystal structure prediction procedures. 

Among the density functional approximations, hybrid functionals are generally more accurate than functionals solely based on the generalized gradient approximation (GGA) --- but also significantly more expensive. 
Several examples indicate that for instance the hybrid PBE0+MBD approach can yield more accurate results for molecular crystals and improve upon the PBE+MBD description at the GGA level\cite{Reilly2013,Reilly2013b,Marom2013,Reilly2015,Shtukenberg2017,Hoja2018,Hoja2019}. 
However, fully converged periodic hybrid calculations can easily become prohibitively expensive for practically relevant systems or in cases when a huge number of calculations are required, for instance during crystal structure predictions.
In addition to the immense increase in CPU time compared to GGAs, also memory requirements can often not be satisfied for large unit-cell sizes. 

One possible solution to this problem is the usage of embedding schemes, which approximate the periodic hybrid DFT calculation with less expensive calculations.  
Such embedding approaches typically make use of a molecular many-body expansion, \emph{i.e.}, they involve monomers, molecular dimers, trimers, etc., and are often also referred to as fragment methods\cite{Wen2012,Beran2016,SchmittMonreal2020,Paulus2006,Hermann2009,Bates2011,Bates2011b,Pruitt2012,Herbert2019,Liu2019}. 
Any periodic high-level method can be approximated by either an additive or a subtractive scheme. In the additive case\cite{Yang2014,Cervinka2016,Teuteberg2019}, monomer energies, dimer interactions, trimer interactions, etc. are summed up, eventually converging to the periodic result and thereby completely circumventing an explicitly periodic calculation. 
In contrast, a subtractive scheme\cite{Beran2010,Wen2011,Nanda2012,Boese2017,Loboda2018,Dolgonos2018} involves an explicit periodic calculation at a lower-level method followed by replacing monomer energies, dimer interactions, trimer interactions, etc. with the values from the high-level method. 

Recently, several such embedding approaches have been developed for molecular crystals utilizing even up to MP2 or CCSD(T) as high-level method in a subtractive scheme\cite{Beran2010,Wen2011,Nanda2012,Boese2017,Greenwell2020,Borca2023}. 
Also, Chen and Xu\cite{Chen2020} have introduced a different fragmentation scheme involving only parts of a molecule. 
Instead of including the most expensive methods, embedding is also utilized to approximate for instance GGA calculations by GGA fragments and periodic density-functional tight binding calculations in order to enable very large calculations\cite{Dolgonos2018}. 

One of us has introduced a subtractive embedding scheme for approximating hybrid density functionals\cite{Loboda2018}, which consists of a periodic GGA calculation and a monomer and dimer correction utilizing the hybrid functional. 
This methodology was implemented in Ref. \citenum{Loboda2018} for energies and lattice relaxations and the following hybrid:GGA combinations were tested: 
PBE0:PBE+D3, PBE0:PBE+MBD, and B3LYP:BLYP+D3.

In this paper, we present an extension and a new open-source implementation of the above-mentioned embedding approach in order to speed up or enable hybrid calculations for larger molecular crystals.
Specifically, we extend the energy calculation up to trimers and enable harmonic phonon calculations, which can now be performed utilizing up to dimers.  
We test the performance of the resulting multimer embedding approach by embedding PBE0+MBD\cite{Adamo1999,Tkatchenko2012,Ambrosetti2014} multimers into periodic PBE+MBD\cite{Perdew1996} calculations and comparing with explicit periodic PBE0+MBD results utilizing the X23\cite{Reilly2013,Reilly2013b,OterodelaRoza2012,Dolgonos2019} benchmark set, which has been extensively used to test and develop methods for molecular crystals\cite{Moellmann2014,Ikabata2014,Grimme2015,Cutini2016,Nyman2016,Gould2016,Mortazavi2018,Stein2019,Chen2020,Gale2021,Jana2021,Tuca2022,Price2023}.

\section{Computational Methods}

\subsection{Energy}

Within our subtractive multimer embedding scheme, the periodic high-level energy $E_\mathrm{per}^\mathrm{high}$ is approximated according to
\begin{equation}
\begin{split}
     \label{eq:energy}
    E_\mathrm{per}^\mathrm{high} \approx E_\mathrm{per}^\mathrm{low} 
    + \sum_i n_i \Delta E_i
    + \sum_{i>j} \frac{n_{ij}}{2} \Delta E_{ij}^\mathrm{int} \\
    + \sum_{i>j>k} \frac{n_{ijk}}{3} \Delta E_{ijk}^\mathrm{int}
\end{split}
\end{equation}
The first term $E_\mathrm{per}^\mathrm{low}$ refers to the fully periodic calculation utilizing a computationally more efficient lower-level method, while the following terms then replace certain energies with results from a high-level method and are explained in detail below. 

Note that we prefer to utilize the term \emph{multimer embedding} in order to indicate that our fragments are complete molecules and in order to refrain from the term \emph{many-body expansion} for specifying multimers, since this may cause confusion with the used many-body dispersion (MBD) method, wherein a body refers to an atom.
We label embedding results ME$X$(PBE0+MBD:PBE+MBD), where the first term in parenthesis is the high-level method, the second term the low-level method, and the $X$ refers to the utilized multimer order, \emph{i.e.}, 1 if only monomers are included, 2 when up to dimers are considered, and 3 when up to trimers are used. 
Dimers and trimers are considered if their shortest intermolecular distance is smaller than a defined multimer cutoff distance.
Currently, for trimers to be considered, the distance between all pairs of molecules within the trimer must have an intermolecular distance smaller than the multimer cutoff value. 
Since we will only utilize PBE0+MBD embedded into PBE+MBD in this paper, we will omit the information in parenthesis but rather use this way to specify the used cutoff distance.

From the input unit cell a supercell of sufficient size is created based on the given multimer cutoff distance, from which then all necessary multimers are extracted.
The sums in eq. \ref{eq:energy} always run over the monomers within the created supercell and the respective $n$ amounts to the number of monomers of a given multimer that belong to the central unit cell. 
For the first sum, the value of $n_i$ is simply 1 in case monomer $i$ belongs to the central unit cell and 0 otherwise. 
For the second term, $n_{ij}$ is 2 if both monomer $i$ and monomer $j$ belong to the central unit cell,  1 if only monomer $i$ or $j$ belongs to the central unit cell, and 0 otherwise. 
The same is true for the third sum containing trimers.
Any $\Delta E$ term always refers to the difference between the high-level and the low-level quantity
\begin{equation}
    \Delta E = E_\mathrm{high} - E_\mathrm{low}.
\end{equation}

The term $E_{ij}^\mathrm{int}$ corresponds to the dimer interaction energy
\begin{equation}
    E_{ij}^\mathrm{int} = E_{ij} - E_{i} - E_{j} \, ,
\end{equation}
where $ij$ is the dimer consisting of monomers $i$ and $j$.
Similarly, the term $E_{ijk}^\mathrm{int}$ refers to the trimer interaction energy
\begin{equation}
    E_{ijk}^\mathrm{int} = E_{ijk} - E_{ij}^\mathrm{int} - E_{ik}^\mathrm{int} - E_{kj}^\mathrm{int} - E_{i} - E_{j} - E_{k},
\end{equation}
where $ijk$ refers to the trimer consisting of monomers $i$, $j$, and $k$. 
The lattice energy $E^{\rm high}_{\rm latt}$ is then calculated according to 
\begin{equation}
    E^{\rm high}_{\rm latt} = \frac{E_\mathrm{per}^\mathrm{high}}{Z} - E_\mathrm{mon}^\mathrm{high},
\end{equation}
where $Z$ is the number of molecules within the unit cell and $E_\mathrm{mon}^\mathrm{high}$ is the high-level energy of one monomer in its most stable gas-phase conformation.

\subsection{Forces}

For any given atom $a$ belonging to monomer $i$ within the central unit cell, the approximated high-level force acting on it is given by
\begin{equation}
    \mathbf{f}_\mathrm{per}^\mathrm{high}(a) \approx \mathbf{f}_\mathrm{per}^\mathrm{low}(a)
    +  \Delta \mathbf{f}_i(a)
    + \sum_{j} \Delta \mathbf{f}_{ij}^\mathrm{int}(a) \, ,
\end{equation}
with $\mathbf{f}_\mathrm{per}^\mathrm{low}(a)$ being the force on atom $a$ from the low-level periodic calculation; $\Delta \mathbf{f}_i(a)$ refers to the difference in the forces on atom $a$ between the high-level and low-level calculation of monomer $i$, and $\mathbf{f}_{ij}^\mathrm{int}(a)$
 is the dimer interaction force at atom $a$ (with $i\neq j$)
 \begin{equation}
    \mathbf{f}_{ij}^\mathrm{int}(a) = \mathbf{f}_{ij}(a) - \mathbf{f}_{i}(a).
\end{equation}

\subsection{Stress}

The stress tensor $\mathbf{\sigma}$ is a second-order symmetrical tensor consisting of nine components
\begin{equation}
    \mathbf{\sigma} = 
    \begin{pmatrix}
        \sigma_{11} & \sigma_{12} & \sigma_{13} \\
        \sigma_{21} & \sigma_{22} & \sigma_{23} \\
        \sigma_{31} & \sigma_{32} & \sigma_{33}
    \end{pmatrix},
\end{equation}
six of which are unique. 
Expressions for the derivative of lattice parameters or the stress tensor for fragment/multimer methods have for instance been published by Nanda and Beran\cite{Nanda2012} and by Loboda et al.\cite{Loboda2018}. Here, we approximate the stress tensor components of our high-level method with
\begin{equation}
\begin{split}
    \sigma_{pq}^\mathrm{high} \approx \sigma_{pq}^\mathrm{low} 
    -\frac{1}{V} 
    \sum_i  \sum_{a} n_i r_{i,p}(a)\, \Delta f_{i,q}(a)\\
    -  \frac{1}{V}  \sum_{i>j}  \sum_{a} \frac{n_{ij}}{2} r_{ij,p}(a) \, \Delta f_{ij,q}^\mathrm{int}(a),
\end{split}
\end{equation}
where $\sigma_{pq}^\mathrm{low}$ are the obtained stress tensor components from the low-level periodic calculations and $V$ is the unit cell volume. The first summation in both terms sums up over all monomers and dimers, respectively. 
The second summation runs over all atoms $a$ in the respective multimer. 
The meaning of $n$ is the same as above and accounts for the number of  monomers within the central unit cell.  
The indices $p$ and $q$ range from 1 to 3 and indicate weather the $x$, $y$, or $z$ component of the position $\mathbf{r}$ and force $\mathbf{f}$ of atom $a$ is to be used.

\subsection{Harmonic Vibrational Properties}

For the calculation of vibrational properties, we utilize phonopy\cite{Togo2015} to create the necessary finite displacements within sufficiently large supercells. 
The corresponding high-level atomic forces are approximated as described in section B, but this time applying the correction to the whole phonopy supercell.
So for every displaced atom, all corresponding displaced multimers are calculated and the periodic force constants corrected accordingly.
With the approximated high-level force sets, the vibrational properties are then calculated within phonopy.

\subsection{Computational Details}

All shown multimer embedding calculations were performed by using our new open-source code MEmbed\cite{Membed,membed2} (version 0.2.0). 
All electronic structure calculations were performed by utilizing FHI-aims\cite{Blum2009,Knuth2015,Ren2012,Yu2018,Havu2009,Ihrig2015} (version 210716\_2), which enables the calculation of isolated and periodic systems on an equal footing due to the employed numeric atom-centered basis functions. 
MEmbed makes use of the Atomic Simulation Environment\cite{HjorthLarsen2017} (ASE) for several of its functionalities, including an interface with FHI-aims. 

Throughout, we used either the PBE\cite{Perdew1996} or the PBE0\cite{Adamo1999} density functional approximation together with the many-body dispersion\cite{Tkatchenko2012,Ambrosetti2014} (MBD) method (rsSCS version) for proper accounting of van-der-Waals dispersion interactions. 
Most calculations were performed by utilizing the light species default settings within FHI-aims for integration grids and basis functions (version 2020) in order to allow for the calculation of the canonical periodic PBE0+MBD method as reference for the multimer embedding approach. 
Performing all calculations in a fully periodic way with PBE0+MBD using converged tight settings would not be feasible due to the massive amount of required CPU time and memory, especially for the supercells needed for the phonon calculations.
However, in order to validate the multimer embedding approach also for tight settings, we have additionally performed single-point energy calculations and lattice relaxations with tight settings (version 2020 defaults plus one additional auxiliary g function to improve the resolution of identity approximation\cite{Ihrig2015}) but lattice relaxations of the canonical PBE0+MBD/tight method were restricted to a subset of 8 small structures from X23.

Note that these embedding calculations differ from the corresponding results by Loboda et al.\cite{Loboda2018}. 
Therein, periodic calculations were performed using plane waves and pseudopotentials in VASP, while the isolated multimers were calculated with a TZVPPD\cite{Weigend2005,Rappoport2010} Gaussian basis set within TURBOMOLE. This lead to some inconsistency since in this case the low-level description of the periodic system and the multimers is not completely identical.
Furthermore, all dispersion corrections were calculated using the PBE0 range-separation parameter for MBD\cite{Loboda2018}, effectively using the multimer scheme only for the DFT part. 
Herein, by using FHI-aims, we are able to perform all calculations (periodic and isolated multimers) on a completely equal footing in terms of software and basis sets since we utilize all-electron calculations with numeric atom-centered basis functions.
In addition, we use for all PBE and PBE0 calculations MBD with the respective default value for the range-separation parameter. 

In all FHI-aims single-point calculations the total energy, the forces, the charge density, and the sum of eigenvalues were converged to $10^{-6}$~eV, $10^{-4}$~eV/\AA{}, $10^{-5}$~electrons/\AA{}$^3$, and $10^{-5}$~eV, respectively. 
The k-grids for the periodic DFT calculations were set to always satisfy $n x > 18$~\AA{}, with $x$ being the cell length in the respective direction and $n$ the number of k-points in that direction.
For MBD energies and forces a tighter k-grid satisfying $n x > 25$~\AA{} was used. 
All phonon calculations were performed utilizing finite displacements of 0.005~\AA{} and appropriate large supercells were created so that the length in every direction exceeds 12~\AA{}. The q-grid used for the evaluation of the vibrational free energy was set to satisfy $n x > 50$~\AA{}. 
This resulted in all acoustic modes being smaller than 0.8~cm$^{-1}$ in magnitude at the gamma point. 
All lattice relaxations were performed until a convergence of 0.005~eV/\AA{} was reached using ASE with the BFGS algorithm and the ExpCellFilter\cite{Tadmor1999} class, as well as the FixSymmetry class to maintain symmetry. 
For the calculation of lattice energies, we utilized for the isolated monomers the structures provided in Ref. \citenum{Reilly2013} as a starting point and re-optimized them with the respective methods using a convergence criteria of 0.001~eV/\AA{}.

\section{Results and Discussion}

\subsection{Embedding for Lattice Energies}

First, we assess the performance of our multimer embedding approach for lattice energies. 
In order to eliminate any geometry effects within this comparison, we compare only single point calculations carried out with different approaches on top of the PBE0+MBD/light-optimized structures. 
The systems of the X23 benchmark set are listed together with their number of molecules within the unit cell ($Z$) and their number of atoms per molecule ($n$) in Table S1 in the Supporting Information (SI). 
Among the X23 systems $n$ varies between 3 atoms (carbon dioxide) and 26 atoms (adamantane), while $Z$ varies between 2 and 8, yielding unit cells containing between 12 and 72 atoms.

In order to illustrate the complexity of the evaluated embedding approaches, Table S1 also contains for each multimer cutoff distance the number of identified unique dimers and trimers. 
Note that, whenever possible, symmetry constraints are utilized in order to limit the number of unique multimers to calculate. 
For instance, without symmetry a 3~\AA{} cutoff ME3 calculation for the ammonia crystal would already require 42 dimer and 76 trimer calculations. 
In contrast, with symmetry only 2 dimers and 7 trimers are necessary in this case.
For a succinic acid crystal, which involves slightly larger and less symmetric molecules with 14 atoms, symmetry can also significantly reduce the costs of a ME3 calculation with 3~\AA{} cutoff from 19 dimers and 22 trimers to only 5 dimers and 4 trimers.\\

\begin{figure*}
    \centering
    \includegraphics[width=\textwidth]{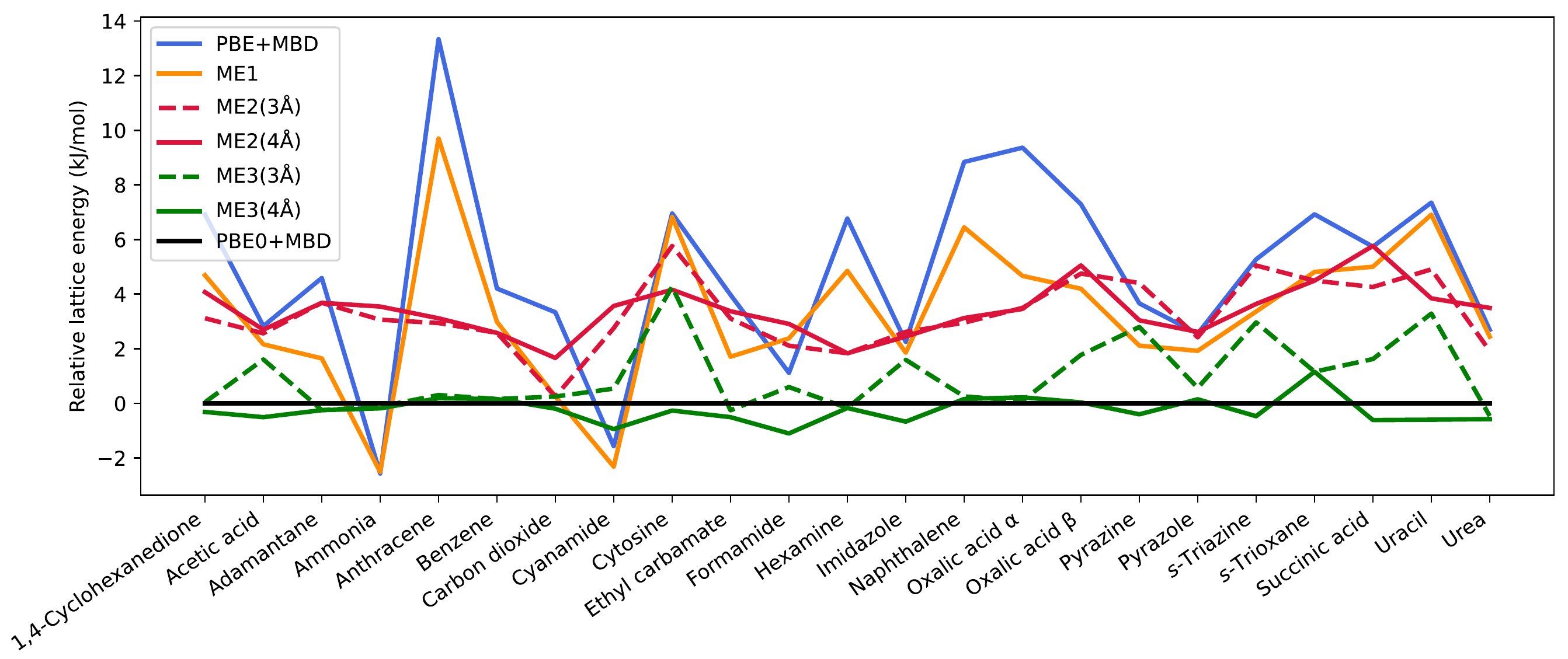}
    \caption{Relative lattice energies of the X23 set for several approaches w.r.t. PBE0+MBD in kJ/mol.}
    \label{fig:energies}
\end{figure*}
\begin{table}
\setlength{\tabcolsep}{2pt}
\caption{\label{tbl:energies} Errors of the calculated lattice energies of the X23 set compared to PBE0+MBD. All calculations were done with light settings on top of the PBE0+MBD-optimized structures. The mean error (ME), the mean absolute error (MAE), and the maximal error (MAX) are given in kJ/mol while the mean relative error (MRE), the mean absolute relative error (MARE), and the maximal relative error (RMAX) are given in \%. }
\begin{tabular}{@{\extracolsep{4pt}}lrrrrrr@{}}
\hline\hline
Method &       ME & MAE & MAX  & MRE    & MARE & RMAX\\
\hline
PBE+MBD     & 4.9 & 5.2 & 13.3 &  -5.2  &   5.9 & 11.9 \\
ME1         & 3.3 & 3.7 &  9.7 &  -3.2  &   3.9 & 7.8  \\
ME2(3\AA{}) & 3.3 & 3.3 &  5.8 &  -3.7  &   3.7 & 7.9  \\
ME2(4\AA{}) & 3.4 & 3.4 &  5.8 &  -4.0  &   4.0 & 8.3  \\
ME2(5\AA{}) & 3.4 & 3.4 &  5.6 &  -4.0  &   4.0 & 8.3  \\
ME2(6\AA{}) & 3.5 & 3.5 &  5.6 &  -4.2  &   4.2 & 9.3  \\
ME2(7\AA{}) & 3.5 & 3.5 &  5.5 &  -4.1  &   4.1 & 9.0  \\
ME2(8\AA{}) & 3.5 & 3.5 &  5.5 &  -4.2  &   4.2 & 9.0  \\
ME3(3\AA{}) & 1.0 & 1.1 &  4.3 &  -1.0  &   1.2 & 4.6  \\
ME3(4\AA{}) &-0.2 & 0.4 &  1.2 &   0.3  &   0.5 & 1.6  \\
ME3(5\AA{}) &-0.6 & 0.6 &  1.5 &   0.7  &   0.7 & 2.1  \\
ME3(6\AA{}) &-0.8 & 0.8 &  2.1 &   0.9  &   1.0 & 2.7  \\
\hline\hline
\end{tabular}\\
\end{table}

We start by first discussing lattice energies obtained with light species default settings since these settings can be used to evaluate the performance of our multimer embedding approach all the way up to harmonic vibrational properties due to the fact that corresponding canonical periodic PBE0+MBD calculations are still computationally feasible. 
Fig. \ref{fig:energies} shows relative lattice energies of our low-level method (PBE+MBD/light) and a selection of embedding approaches for all X23 systems compared with the used high-level method (PBE0+MBD/light) and Table \ref{tbl:energies} lists several error statistics for all considered approaches.
Since the lattice energies have a negative sign, positive values in the plot and the reported mean errors (ME) indicate a smaller interaction magnitude than the reference. 
The corresponding lattice energies of all systems are listed in Tables S2 and S3 in the SI. 
All structures and detailed calculation results including all energies of all isolated multimers are further available in a Zenodo repository\cite{sidata} as ASE database files.

It can be seen that PBE+MBD leads for almost all systems to an underbinding with a ME of 4.9~kJ/mol and a mean relative error (MRE) of -5.2~\% compared to PBE0+MBD; only for ammonia and cyanamide PBE+MBD yields a larger lattice energy in magnitude. 
Hence, the mean absolute error (MAE) is slightly larger than the ME, amounting to 5.2~kJ/mol or 5.9~\% in terms of the mean absolute relative error (MARE). 
In the worst case (anthracene), the largest observed absolute difference between PBE+MBD and PBE0+MBD is 13.3~kJ/mol.

The monomer embedding (ME1) shows the same qualitative trend as PBE+MBD. 
However, the inclusion of only monomers at the PBE0+MBD level already reduces the ME to 3.3~kJ/mol as well as the MAE to 3.7~kJ/mol, which accounts for about 30~\% of the difference between PBE0+MBD and PBE+MBD.   
When dimers are included, all lattice energies are now smaller in magnitude than the reference and at a multimer cutoff of 3~\AA{} the MAE is reduced to 3.3~kJ/mol.
When utilizing larger cutoff values the errors remain quite similar but actually increase slightly. 

As soon as trimers are included, we see a significant improvement in terms of all errors and now some lattice energies become again larger in magnitude than the reference value.
By only considering dimers and trimers up to 3~\AA{} the ME and MAE can be reduced to 1.0 and 1.1~kJ/mol, respectively. 
When moving on to 4~\AA{}, we observe the best agreement with PBE0+MBD with the ME, MAE, and the maximal error being only -0.2, 0.4, and 1.2~kJ/mol, respectively. 
At this level the high-level lattice energy is extremely well approximated when considering fixed geometries. 
Increasing the multimer cutoff further leads to a small increase in all errors. 
 
In terms of convergence of the lattice energies with the used multimer order, a part of the systems always improve with increasing multimer order, while others basically follow a damped oscillation. The latter can be seen for instance for ammonia. There, the addition of dimers leads to a quite substantial underbinding, resulting in larger errors as at the ME1 level. Adding trimers leads to a smaller correction in the opposite direction, which significantly reduces the error. In general, utilizing trimers with a larger multimer cutoff leads in this case for X23 to a small overbinding.
So in order to further reduce the errors at large multimer cutoffs, tetramer energies would probably need to be included. However, there seems to be a quite beneficial error cancellation at small multimer cutoffs, so that 3 or 4~\AA{} are sufficient here. 

After having established the convergence behavior of our multimer embedding approach with light species default settings, we discuss now lattice energies obtained with converged tight species default settings, which were calculated on top of PBE0+MBD/light-optimized structures. 
The obtained individual lattice energies are listed in Table S4 in the SI and the corresponding statistical errors w.r.t. periodic PBE0+MBD are given in Table S5 in the SI.
For tight settings the difference between PBE0+MBD and PBE+MBD is for X23 smaller than for the light settings with a MAE of 3.0~kJ/mol. 
Monomer embedding (ME1) reduces the MAE to 2.2~kJ/mol. 
Adding now dimer corrections actually increases the MAE to for instance 2.8~kJ/mol at a cutoff of 3~\AA{}.  
As soon as trimer interactions are corrected, the errors drop significantly --- just like in the case of light settings. Here, we even reach already at the ME3(3\AA{}) level a MAE below 1~kJ/mol (0.7~kJ/mol) and at the ME3(4\AA{}) level the convergence is very similar to the one with light settings, with now ME, MAE, and MAX values of only -0.5, 0.5, and 1.4~kJ/mol, respectively.

\subsection{Embedding for Forces and Stress}

After having evaluated the energies at fixed geometries, we now study forces and stress tensors.
First, we have calculated the atomic forces and the stress tensors at PBE+MBD-optimized structures, so that the components are non-zero for our PBE0+MBD reference method, while they are virtually zero for our utilized low-level method.
The resulting errors are shown in Table \ref{tbl:forces}.

\begin{table}
\setlength{\tabcolsep}{2pt}
\caption{\label{tbl:forces} Mean absolute errors (MAE) and maximal errors (MAX)  of the calculated atomic force components and non-zero stress tensor components of the X23 set compared to PBE0+MBD calculated at PBE+MBD-optimized structures.  }
\begin{tabular}{@{\extracolsep{4pt}}lrrrr@{}}
\hline\hline
 &\multicolumn{2}{c}{Forces (eV/\AA{})} & \multicolumn{2}{c}{Stress (eV/\AA{}$^3)$} \\  \cline{2-3}\cline{4-5}
Method &       MAE & MAX  &  MAE & MAX\\
\hline
PBE+MBD     & 0.226 & 1.422 & 0.0190 & 0.0496\\
ME1         & 0.023 & 0.216 & 0.0009 & 0.0034\\
ME2(3\AA{}) & 0.008 & 0.120 & 0.0007 & 0.0039\\
ME2(4\AA{}) & 0.006 & 0.055 & 0.0007 & 0.0039\\
ME2(5\AA{}) & 0.006 & 0.057 & 0.0008 & 0.0039\\
ME2(6\AA{}) & 0.006 & 0.048 & 0.0008 & 0.0040\\
\hline\hline
\end{tabular}\\
\end{table}

It can be seen that this resulted in an average absolute force component difference between PBE0+MBD and PBE+MBD of about 0.2~eV/\AA{}. 
The monomer embedding already significantly reduces the MAE by a factor of ten to 0.023~eV/\AA{} and with dimer embedding, it is further reduced by another factor of four to 0.006~eV/\AA{}, which is close to our optimization convergence criterion. 
This implies that the atomic force contributions are already well approximated by dimer embedding. 

In terms of the non-zero components of the stress tensor, we are able to reproduce them with a MAE of 7 $\times$ 10$^{-4}$~eV/\AA{} in the case of ME2(3\AA{}) and ME2(4\AA{}).
Note that the dimers have a small effect on stress tensor components within the currently used approximation. 
While these errors might look very promising, they are still large enough to lead to quite different cell volumes for systems with flat potential energy surfaces like molecular crystals, as we will discuss below. 

Next, we have performed lattice relaxations utilizing monomer and dimer embedding up to 5~\AA{} with light settings. The errors of the resulting cell volumes and the corresponding lattice energies compared to the optimized PBE0+MBD/light values are listed in Table \ref{tbl:opt} and the individual volumes can be found in Table S6 in the SI.

\begin{table}
\setlength{\tabcolsep}{2pt}
\caption{\label{tbl:opt}Errors of calculated cell volumes (in \% ) and corresponding lattice energies (in kJ/mol) of the X23 set calculated with light settings compared to PBE0+MBD results.  }
\begin{tabular}{@{\extracolsep{4pt}}lrrrrrr@{}}
\hline\hline
 &\multicolumn{3}{c}{$V$ (\%)} & \multicolumn{3}{c}{$E_{\rm latt}$ (kJ/mol)} \\  \cline{2-4}\cline{5-7}
Method &    MRE & MARE & RMAX &   ME &  MAE & MAX\\
\hline
PBE+MBD     & 3.8 & 3.8 & 6.3 &  1.7 & 2.4 & 9.2\\
ME1         & 2.9 & 2.9 & 5.5 &  3.2 & 3.7 & 9.9\\
ME2(3\AA{}) & 2.4 & 2.4 & 5.0 &  3.2 & 3.2 & 5.8\\
ME2(4\AA{}) & 2.3 & 2.3 & 5.4 &  3.4 & 3.4 & 5.9\\
ME2(5\AA{}) & 2.4 & 2.4 & 6.5 &  3.5 & 3.5 & 5.9\\
\hline\hline
\end{tabular}\\
\end{table}

PBE+MBD/light overestimates the X23 cell volumes compared to PBE0+MBD/light with a mean relative error (MRE) of 3.8~\%. 
All shown embedding approaches also always overestimate the cell volume compared to PBE0+MBD. 
In case of monomer embedding, the mean absolute relative error (MARE) can be reduced to 2.9~\% and the best dimer embedding (4\AA{}) leads to a MARE of 2.3~\%. 
Hence, the accuracy of the stress tensor is not yet sufficient at the dimer level to accurately approximate the PBE0+MBD lattice constants. 

Table \ref{tbl:opt} also shows the errors of the resulting lattice energies at the optimized cells when compared to those of canonical PBE0+MBD lattice relaxations. 
These errors are very similar to the results for the frozen PBE0+MBD structures. 
This indicates that the potential energy surfaces of these systems are indeed very flat and that a very high accuracy of the stress tensor is probably needed to actually reproduce the PBE0+MBD cell volumes. 
Interestingly, the PBE+MBD mean errors are now much smaller than for the frozen PBE0+MBD structures, which is due to the fact that now at the actual PBE+MBD equilibrium structures the resulting lattice energies have increased in magnitude.

\begin{figure}
    \centering
    \includegraphics[width=\columnwidth]{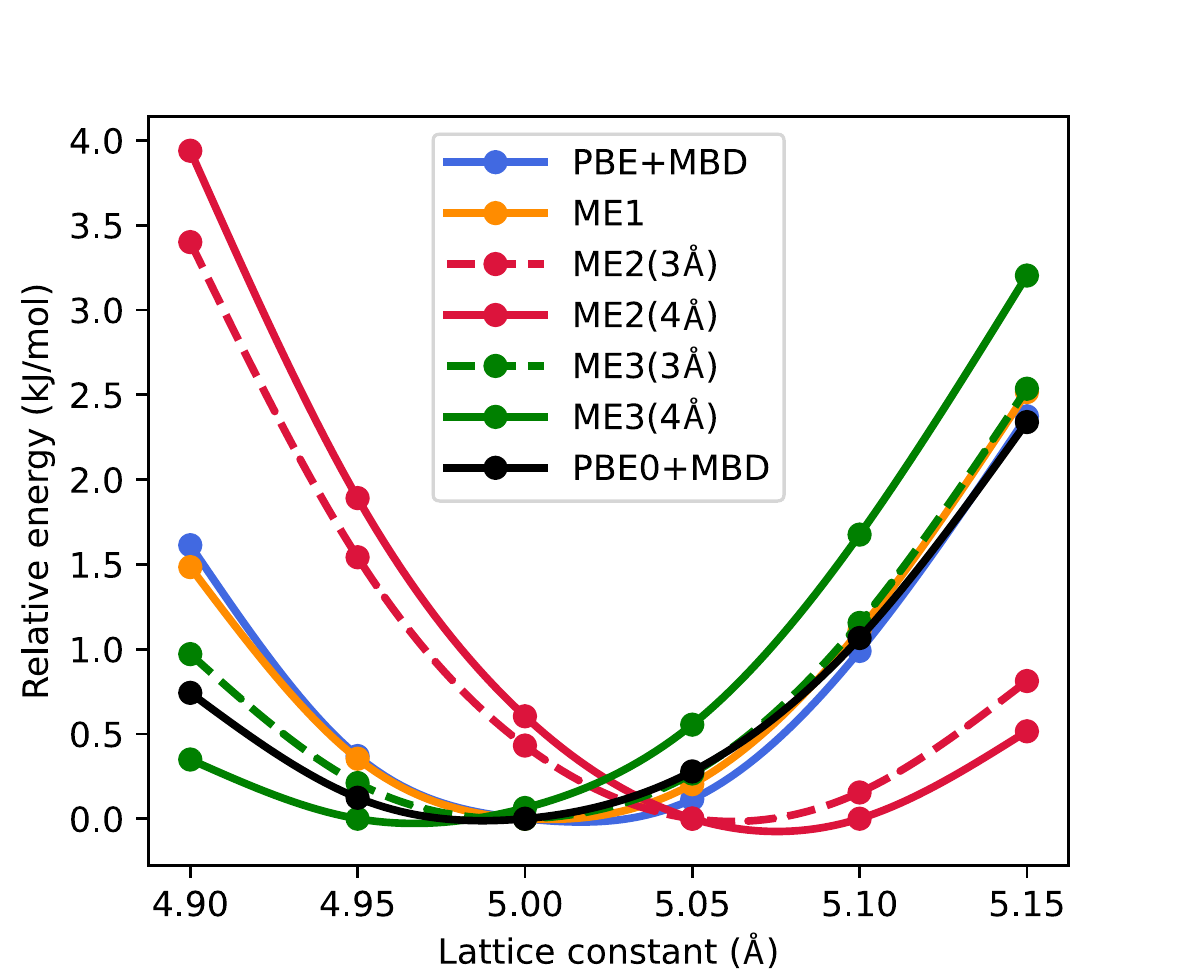}
    \caption{Relative energies of a cubic ammonia unit cell w.r.t the lattice constant for several embedding approaches (light settings). }
    \label{fig:nh3}
\end{figure}

In order to illustrate the impact of trimer interactions on lattice constants and unit-cell volumes, we have calculated as example single-point energies of cubic ammonia at varying lattice constants (see Fig. \ref{fig:nh3}). 
Our dimer embedding significantly overestimates the lattice constant and hence the cell volume compared to PBE0+MBD as well as to PBE+MBD for the ammonia crystal. 
In fact, ammonia is for all cutoffs at the ME2 level the system with the worst agreement with the periodic PBE0+MBD results in terms of the cell volume. 
However, we note that the optimal unit-cell volumes for ME2(3\AA{}) and ME2(4\AA{}) obtained via a Murnaghan equation-of-state\cite{Murnaghan1944} fit from the single-point energies in Fig. \ref{fig:nh3} respectively agree within 0.3~\% and 0.1~\% with the corresponding embedding optimizations, which further validates the used stress-tensor approximation.

When trimers are included, the lattice constant gets significantly smaller. 
At the ME3(3\AA{}) level there is an excellent agreement with the PBE0+MBD value; moving to larger multimer cutoffs seems to lead to a slight underestimation of the lattice constant. 
This illustrates that utilizing trimers within lattice relaxation could indeed significantly improve our description of lattice constants.
Since the number of trimers to be considered can be quite large and is increasing heavily with increasing cutoff, an efficient inclusion of trimer interactions for the calculation of forces and stress tensors requires an in-depth study of which trimers are important and which could be omitted. 
Therefore, we will discuss explicit lattice relaxations with trimer interactions in a follow-up publication investigating also the performance of different dimer/trimer cutoff combinations and exploring other ways to reduce the number of considered trimers.

Next, we have also performed lattice relaxations utilizing converged tight species default settings. 
Given the massive computation time of the canonical periodic PBE0+MBD/tight calculations, we only compare a subset of X23 containing 8 structures, for which the reference calculations were still computationally feasible. 
The resulting individual unit-cell volumes are given in Table S7 and the errors compared to PBE0+MBD/tight in Table S8 in the SI. 
In this case, PBE+MBD/tight has a MARE of 2.5~\% for the cell volume and both monomer embedding as well as dimer embedding reduce the error to 1.8~\%. 
The convergence behavior for dimer embedding is very similar between light and tight settings; in both cases the PBE+MBD MARE of the cell volumes is 33~\% larger than the corresponding ME2(4\AA{}) MARE.

So far, we have only evaluated the performance of the multimer embedding by comparison with the canonical PBE0+MBD calculations. 
To put the results a bit more into perspective, we briefly also mention the performance when comparing to reference values derived from experimental sublimation enthalpies. 
In order to directly evaluate static lattice energies, sublimation enthalpies can be back-corrected for vibrational contributions\cite{Reilly2013,OterodelaRoza2012}. 
Here, we utilize our recently introduced X23b reference data\cite{Dolgonos2019}, which also includes a back-correction of experimental volumes in terms of the average thermal expansion of three density functionals, so that the results of lattice relaxations can directly be compared. 
Our PBE0+MBD/light cell volumes have a MARE of 2.4~\%, while PBE+MBD/light has a MARE of 5.9~\%, and with ME2(4\AA{})/light we obtain a MARE of 4.4~\%.
In comparison, the PBE0:PBE+MBD approach by Loboda et al.\cite{Loboda2018} reaches an accuracy of 3.6~\% compared to the X23b reference. 
However, the two approaches are not directly comparable since in this work, we are utilizing light settings for numeric atom-centered basis functions, no pseudopotentials, and standard range-separation parameters for MBD in all calculations.

In terms of lattice energies, our ME2(4\AA{}) approach utilizing light settings reaches a MAE of 4.9~kJ/mol. 
We note that while geometries are often already well described at the light level, it is by far not sufficient for obtaining converged values for energetics.
Utilizing converged tight species default settings the MAE w.r.t. X23b lattice energies amounts for PBE+MBD (optimized) to 3.8~kJ/mol and for PBE0+MBD (calculated on top of PBE0+MBD/light structures) to 3.2~kJ/mol.
At fully optimized ME2(4\AA{})/tight structures this error even decreases to 2.9~kJ/mol, where we seem to benefit from a certain error cancellation.
Small differences in the cell volumes seem to virtually have no effect on the overall accuracy of the corresponding lattice energies when comparing to X23b since the ME2(4\AA{})/tight MAE when using PBE0+MBD/light-optimized structures amounts to also 2.9~kJ/mol.
In comparison, the PBE0:PBE+MBD approach by Loboda et al.\cite{Loboda2018} (when correcting the isolated monomer energies for oxalic acid) reaches an accuracy of 3.6~kJ/mol.

\subsection{Embedding for Harmonic Vibrational Properties}

Next, we evaluate the performance of our embedding approach for vibrations/phonons using the X23 benchmark set. 
Therefore, we have calculated with several methods using light species default settings the gamma-point frequencies for the respective optimized lattice constants and for an internally relaxed structure with the PBE0+MBD lattice constants in order to determine if there are significant changes due to the different lattice constants (see Table \ref{tbl:vib}). 
It can be seen that there are substantial differences between PBE+MBD and PBE0+MBD in terms of vibrational frequencies with a MAE of almost 50~cm$^{-1}$. These large differences originate mainly from the higher-frequency modes; when comparing only the first 300~cm$^{-1}$ the resulting MAEs for the PBE0+MBD cell and for the optimized structure are only 3.3 and 6.6~cm$^{-1}$, respectively.

\begin{table}
\setlength{\tabcolsep}{5pt}
\caption{\label{tbl:vib} Mean errors (ME) and mean absolute errors (MAE) of the gamma-point vibrational/phonon frequencies  of the X23 set  in cm$^{-1}$  and vibrational free energies at 300~K normalized per molecule in kJ/mol (converged q-grid) compared to PBE0+MBD results (light settings). In one case the structures are internally optimized utilizing the PBE0+MBD lattice constants and in the other case the structures are fully optimized. }
\begin{tabular}{@{\extracolsep{2pt}}lrrrrr@{}}
\hline\hline
 & &\multicolumn{2}{c}{PBE0+MBD cell} & \multicolumn{2}{c}{Optimized} \\  \cline{3-4}\cline{5-6}
 Quantity       & Method      &    ME &  MAE &    ME &  MAE \\
\hline
                & PBE+MBD     & -45.5 & 47.1 & -48.2 & 49.0 \\
                & ME1         &   0.9 &  3.0 &  -1.2 &  4.2 \\
$\nu(\Gamma)$   & ME2(3\AA{}) &   0.6 &  1.7 &  -1.1 &  2.5 \\
                & ME2(4\AA{}) &   0.6 &  1.7 &  -1.0 &  2.7 \\
                & ME2(5\AA{}) &   0.5 &  1.8 &  -1.1 &  2.7 \\
\hline
                & PBE+MBD     &  -9.3 &  9.3 & -10.9 & 10.9 \\
                & ME1         &   0.1 &  0.3 &  -1.2 &  1.2 \\
$F_{\rm vib}^a$ & ME2(3\AA{}) &   0.1 &  0.2 &  -0.8 &  0.8 \\
                & ME2(4\AA{}) &   0.1 &  0.2 &  -0.8 &  0.8 \\
                & ME2(5\AA{}) &   0.1 &  0.2 &  -0.8 &  0.8 \\
\hline\hline
\end{tabular}\\
$^a$ evaluated at 300~K and normalized per molecule
\end{table}

\begin{figure*}
    \centering
    \includegraphics[width=0.8\textwidth]{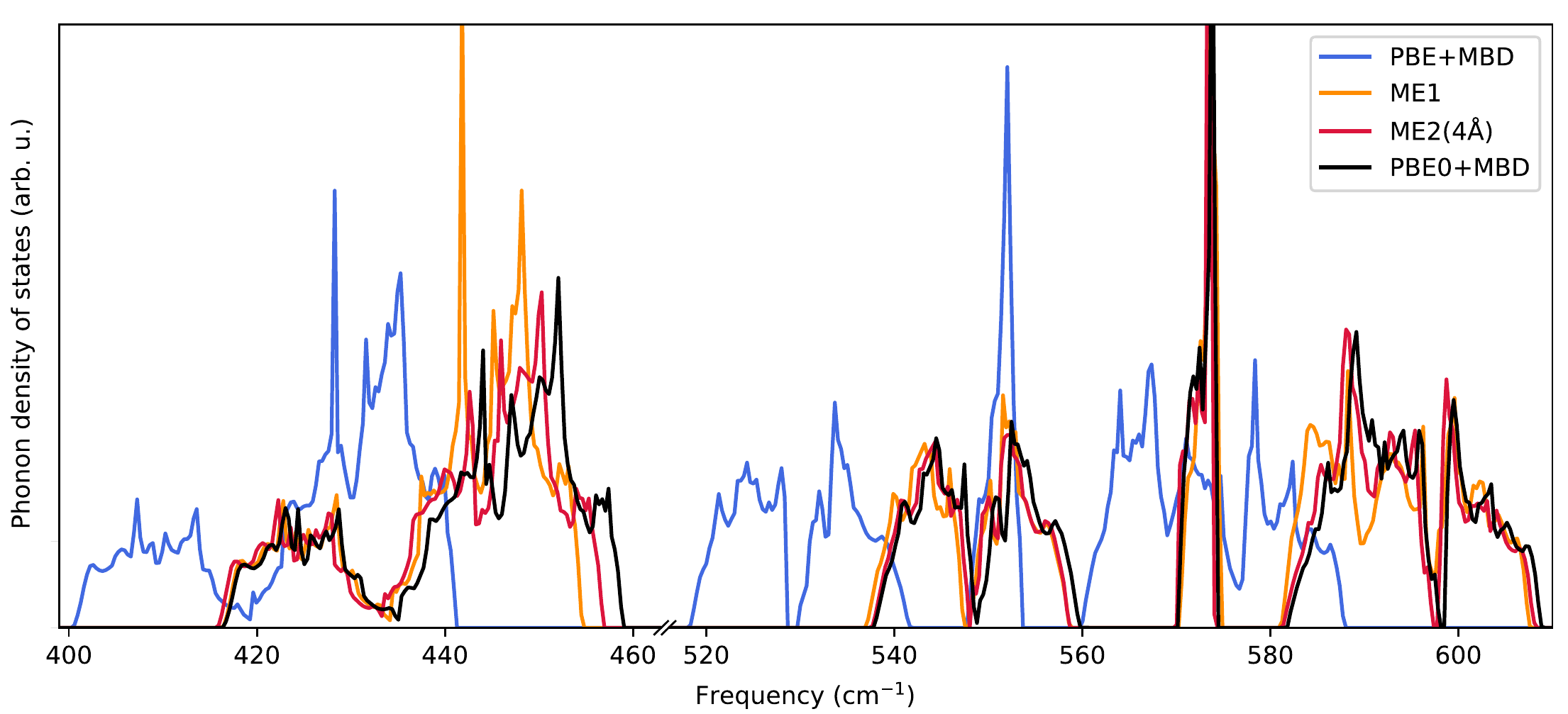}
    \caption{Phonon density of states of uracil calculated on top of optimized structures for several methods (only frequencies between 400 and 610~cm$^{-1}$ are shown).}
    \label{fig:pdos}
\end{figure*}

Monomer embedding already significantly improves the internal vibrational modes and hence produces frequencies with a MAE of about 4~cm$^{-1}$ when evaluated for the entire frequency range. 
When dimers are considered, we can reach an overall MAE of about 3~cm$^{-1}$ and a ME of about 1~cm$^{-1}$ for the optimized structures already at a multimer cutoff of 3~\AA{}.
For the overall statistics, the difference in geometries does not seem to have a large impact, but that is simply due to the large number of internal modes. 
When considering only frequencies up to 300~cm$^{-1}$, we observe slightly larger errors with ME2(3\AA{}) and ME2(4\AA{}) having a MAE of 6 and 4~cm$^{-1}$ at the optimized structures, respectively. 

In order to illustrate the differences in vibrational frequencies, we have plotted in Fig. \ref{fig:pdos} the phonon density of states between 400 and 610~cm$^{-1}$ for completely optimized structures of uracil. 
It can be seen that in this range the peaks for PBE+MBD are shifted quite significantly to lower wave numbers compared to PBE0+MBD. 
Since the main difference in this range comes from intramolecular modes, monomer embedding already corrects for most of the differences, and dimer embedding leads to a small further improvement.
The same plot for the frozen PBE0+MBD cells can be found in Fig. S1 in the SI. 
However, in this frequency range the effect of the slightly different cell parameters is rather small. 
In addition, we have also plotted the low-frequency phonon density of states up to about 160~cm$^{-1}$ for optimized and frozen cells in Figs. S2 and S3 in the SI, respectively. 
It can be seen that in this frequency range the small differences in cell parameters lead to some frequency shifts (Fig. S2) while --- when calculated at the PBE0+MBD cells --- all embedding approaches already nicely match the PBE0+MBD result.

Finally, we discuss the accuracy for vibrational free energies.
In Table \ref{tbl:vib} we compare vibrational free energies evaluated at a converged q-grid at a temperature of 300~K and normalized per molecule with the respective PBE0+MBD result. Note that all thermal properties (in 10~K steps from 0 to 300~K) as well as all data to further post process the results using phonopy are available in ASE databases\cite{sidata}.
It can be seen that PBE+MBD deviates by about 10~kJ/mol from PBE0+MBD and that the monomer embedding already provides a very accurate approximation of the PBE0+MBD vibrational free energy. 
For the optimized structures the error is about 1~kJ/mol and for the PBE0+MBD lattice constants the MAE amounts to only 0.3~kJ/mol. 
Including dimers decreases this error further, leading to a MAE of only 0.8 and 0.2~kJ/mol for the optimized structure and the PBE0+MBD lattice constants, respectively. 
The vibrational free energies at room temperature are already tightly converged at a multimer cutoff of 3~\AA{}, increasing it does not lead to further changes in terms of the MAE. 
Since the vibrational free energy consists of the zero-point vibrational energy (ZPVE) and a thermal contribution, we now analyze the accuracy of the ZPVE for the fully optimized cells. 
For PBE+MBD the MAE amounts to 9.8~kJ/mol (per molecule), suggesting that most of the corresponding vibrational free energy error at room temperature originates in fact from the ZPVE. 
After monomers are corrected (ME1) the ZPVE MAE drops down to 0.3~kJ/mol, which is a quarter of the corresponding vibrational free energy error at room temperature. 
Including dimer corrections then further reduces the ZPVE MAE to 0.2~kJ/mol.

\subsection{Timings}

After having discussed the accuracy of our multimer embedding approach, we illustrate on two examples how much computation time can be saved at a certain embedding level compared to the canonical periodic PBE0+MBD calculation. 
Table \ref{tbl:timings} shows relative timings for an ammonia and a succinic acid crystal. All values are normalized to the CPU time of a PBE+MBD calculation of ammonia with light or tight settings, respectively. A value of 1.0 corresponds for light settings to 0.07 CPU hours and for tight settings to 1.22 CPU hours on Intel Xeon Silver 4214R cores.

\begin{table}
\setlength{\tabcolsep}{3pt}
\caption{Relative timings of single-point energy calculations normalized to PBE+MBD/light or PBE+MBD/tight calculations of ammonia calculated on 4 cores (light) and 24 cores (tight).\label{tbl:timings} }
\begin{tabular}{@{\extracolsep{4pt}}lrrrr@{}}
\hline\hline
 &       \multicolumn{2}{c}{Ammonia} & \multicolumn{2}{c}{Succinic acid} \\\cline{2-3}\cline{4-5}
Method & light & tight & light & tight \\
\hline
PBE+MBD     &  1.0 &    1.0 &  2.7 &   2.7 \\
ME1         &  1.0 &    1.0 &  2.9 &   3.6 \\
ME2(3\AA{}) &  1.2 &    1.2 &  7.9 & 24.8 \\
ME2(4\AA{}) &  1.2 &    1.3 &  9.6 & 33.2 \\
ME3(3\AA{}) &  2.0 &    2.8 & 16.4 &  64.1 \\
ME3(4\AA{}) &  2.6 &    3.8 & 34.1 & 151.3 \\
PBE0+MBD    &  3.4 &   93.3 & 15.3 & 258.6 \\
\hline\hline
\end{tabular}\\
\end{table}

Due to recent advances in the implementation of hybrid density functionals in FHI-aims, the two shown light PBE0+MBD calculations are respectively only 3.4 and 5.7 times more expensive than the corresponding PBE+MBD calculations, which make the light setting extremely useful for comparing with the canonical PBE0+MBD calculations. Given the fast implementation, only a small speedup is possible using light settings and for succinic acid the inclusion of trimer interactions leads already to a similar computation time as the canonical method. However, real computational advantage can be achieved using tight settings, which are typically used for accurate energetics, since in this case PBE0+MBD calculations for the two shown small examples are already almost 100 times more expensive than the PBE+MBD calculations. In addition to the CPU time, larger converged periodic PBE0+MBD calculations are also often prohibitively expensive due to massive memory demands, especially when forces and stress tensors are required. In contrast, multimer embedding does not suffer any real memory issues since the largest hybrid calculation at for instance the ME3 level is an isolated trimer.

In the case of ammonia, which is a highly symmetric crystal with 16 atoms per unit cell and 4 atoms per molecule, tight ME2 calculations are about 70 times faster than the canonical PBE0+MBD approach and even ME3(4\AA{}) is still about 25 times faster. 
When the involved molecules get larger and less symmetric --- like in the case of succinic acid with 14 atoms per molecule --- the speedup is not as massive anymore but ME2(3\AA{}) calculations are still 10 times faster as the canonical methods and even ME3(3\AA{}) is 4 times faster.

\section{Conclusion}
We have introduced trimer interactions and harmonic vibrational properties for a subtractive multimer embedding scheme in order to enable larger calculations for molecular crystals utilizing hybrid density functionals, including a new open-source implementation. Due to the fact that only up to trimers have to be calculated with the high-level method (hybrid functional), this approach is very memory efficient and can also be easily parallelized over multimer calculations. Herein, we approximated periodic PBE0+MBD results by performing periodic calculations using only the more efficient PBE+MBD approach and then introducing the effects of PBE0+MBD by improving  monomer energies, dimer interaction energies, and trimer interaction energies. However, we note that this approach can in principle be used for any combination of methodologies but convergence will most likely be slower when using less compatible methods than PBE0+MBD and PBE+MBD.

The performance of the shown approach was evaluated by directly comparing the multimer embedding results for the X23 benchmark set of molecular crystals with canonical periodic hybrid calculations. In order to accurately approximate lattice energies, the newly incorporated trimer energies are crucial, enabling an agreement within 1~kJ/mol. For lattice relaxations, the dimer embedding yields an error of about 2~\% in terms of the cell volume. A numerical test on the ammonia crystal illustrated that trimer interactions can significantly further improve the description of the cell volume. Hence, the next crucial step towards improving this multimer embedding methodology is the explicit inclusion of trimer interactions for gradients and stress tensors and to reduce the number of multimers that need to be considered to improve the efficiency of this methodology.

Furthermore, we have also newly introduced the calculation of vibrational properties utilizing multimer embedding. We are able to approximate gamma-point vibrational/phonon frequencies with an accuracy of a few wave numbers using monomer or dimer embedding. This enables a very accurate approximation of room temperature vibrational free energies within 1~kJ/mol on average when normalized per molecule in the unit cell. 

This multimer embedding approach at the dimer level can already for single-point energies be up to 70 times faster than the canonical high-level periodic calculation in the case of ammonia when embedding PBE0+MBD into PBE+MBD using converged tight settings within FHI-aims. Since the largest speedup is observed for small monomers, this could potentially be especially relevant for modeling hydrates.

\begin{acknowledgments}
This project has received funding from the European Union’s Horizon 2020 research and innovation programme under the Marie Skłodowska-Curie grant agreement No 890300.
The computational results presented have been achieved in part using the Vienna Scientific Cluster (VSC).
\end{acknowledgments}

\end{document}